\documentclass[final,3p,times,twocolumn]{elsarticle}

\journal{Nuclear Instrument and Method}

\usepackage{graphicx}
\usepackage{amssymb}

\usepackage{lineno}
\begin{document}
\begin{frontmatter}

\title{High Purity Pion Beam at TRIUMF}

\author[label1]{A.Aguilar-Arevalo}
\author[labelvpi]{M. Blecher}
\author[labelubc]{D.A. Bryman}
\author[labelasu]{J. Comfort}
\author[label1]{J. Doornbos}
\author[label1]{L. Doria}
\author[labelnubc]{A. Hussein}
\author[labelosaka]{N. Ito}
\author[labelbnl]{S. Kettell}
\author[label1]{L. Kurchaninov}
\author[labelubc]{C. Malbrunot}
\author[label1]{G.M. Marshall}
\author[label1]{T. Numao\corref{cor1}}
\cortext[cor1]{Corresponding author: E-mail:toshio@triumf.ca, 
Tel:1-604-222-7345}
\author[label1]{R. Poutissou}
\author[label1]{A. Sher}
\author[labelubc]{B. Walker}
\author[labelosaka]{K. Yamada}
\address[label1]{TRIUMF, 4004 Wesbrook Mall, Vancouver,
 B.C. V6T 2A3, Canada}
\address[labelvpi]{Physics Department, Virginia Tech., 
Blacksburg, VA 24061, USA}
\address[labelubc]{Department of Physics and Astronomy, 
University of British Columbia, 
Vancouver, B.C. V6T 1Z1, Canada}
\address[labelasu]{Arizona State University, Tempe, AZ 85287, USA}
\address[labelnubc]{University of Northern British Columbia, 
Prince George, B.C. V2N 4Z9, Canada}
\address[labelosaka]{Physics Department, 
Osaka University, Toyonaka, Osaka, 560-0043, Japan}
\address[labelbnl]{Brookhaven National Laboratory, Upton, NY 11973-5000, USA}

\begin{abstract}

\noindent
An extension of the TRIUMF M13 low-energy pion channel designed
 to suppress positrons based on an energy-loss technique is described.
A source of beam channel momentum calibration from the decay
 $\pi^+ \rightarrow \mbox{e}^+ \nu$ is also described.

\end{abstract}
\end{frontmatter}

\section{Motivation}

The branching ratio of pion decays \cite{triumf,psi},  
R=$\Gamma (\pi \rightarrow \mbox{e} \nu + \mbox{e} \nu \gamma)/
\Gamma (\pi \rightarrow \mu \nu + \mu \nu \gamma)$,
 has provided the
 best test of the hypothesis of electron-muon universality
 in weak interactions.
The new TRIUMF PIENU experiment \cite{pienu} aiming to improve the precision
 of the branching ratio measurement by a factor of five or more
measures positrons from the $\pi^+ \rightarrow \mbox{e}^+ \nu$
decay (E$_{\mbox{e}^+} = 69.8$ MeV)
and the  $\pi^+ \rightarrow \mu^+ \rightarrow \mbox{e}^+$
decay chain ($\pi^+ \rightarrow \mu^+ \nu$ decay followed by
$\mu^+ \rightarrow \mbox{e}^+ \nu \overline{\nu}$ decay, 
E$_{\mbox{e}^+} < 52.8$ MeV).
In order to obtain maximum acceptance with minimum
uncertainties arising from positron energy-dependent cross sections, 
the detector system
involving a large NaI crystal is placed on the beam axis. 
Positrons in the beam (1/4 of the rate
of pions) severely increase detector rates,
trigger rates and background in
the  $\pi^+ \rightarrow \mbox{e}^+ \nu$ spectrum.
The TRIUMF M13 channel \cite{oldm13}  has therefore been upgraded 
to suppress the positron contamination in the pion beam.
\\

\section{ M13 Extension Design}

\begin{figure}[htb]
\centering
\includegraphics*[width=8cm]{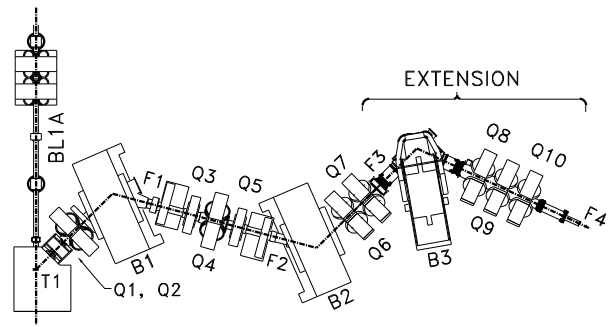}
\caption{M13 channel with the extension.}
\label{beamline}
\end{figure}

\subsection{Existing M13 channel}

The layout of the existing M13 channel \cite{oldm13} together with
 the new beam line extension is shown in Fig.\ref{beamline}.
 The beam line takes off from the primary proton beam line (BL1A) at an
angle of $135^{\circ}$ from a 1-cm thick Be production target (T1).
The M13 channel with a maximum angular acceptance of 29 msr
 is a low-momentum achromatic channel 
with $-60^{\circ}$ (B1 magnet) and
$+60^{\circ}$ (B2 magnet) bends, a quadrupole doublet
(Q1--Q2) between the production target and B1 for collecting pions,
a quadrupole triplet (Q3--Q5) between the two bends,
and a quadrupole doublet (Q6--Q7) downstream of B2 for the final
focusing. There are three foci:
F1 between B1 and Q3, F2 between Q5 and B2, and F3 after Q7.
Beam acceptance-defining slits SL0 are located just upstream of the
first bending magnet B1, and there are momentum-defining
slits SL1 and SL2 at F1 and F2, respectively. About 10 cm downstream 
of SL1, there are two
absorber/slit wheels each with four mounting positions 
that allow $4 \times 4$ combinations of slit 
and/or absorber settings.
The maximum 75-MeV/c pion yield at F3 is 
0.8 M/s for 500 MeV, 100 $\mu$A TRIUMF cyclotron operation.
\\

\subsection{Design Principle}

Energy-loss-based particle separators have been used since the early days of
particle physics experiments \cite{meyer}.
The differential energy loss at 75 MeV/c for pions and positrons
is large enough for clean particle separation, and
this concept can  be applied simply to
the existing M13 structure. When a thin foil is inserted at the absorber
wheel near F1, a momentum spread between pions and positrons
results due to the energy loss difference. 
Following subsequent momentum analysis by B2, pions
and positrons separate into two distinct horizontal distributions at F3.
A collimator can be placed such that it stops only positrons. 
A small momentum tail of the positron beam
due to bremsstrahlung results in some positrons at the pion spot, giving
$<1$ \% positron contamination.
Due to the energy-loss variation in the absorber, the pion beam
has a significant low-momentum tail, which results in 
degradation of the image at F3.
Showers from stopped positrons may be a source of background
with a 23 MHz radio frequency (RF) structure of 
the proton beam from the TRIUMF  cyclotron
if the detector system were located near F3.

The extension of the M13 channel provides controlled pion beam quality,
although there is a pion intensity loss due to 
decay-in-flight.
The blurred pion beam image due to a combination of momentum spread 
by the absorber
and  dispersion by the B2 magnet can be redefined with a collimator at F3
for an improved image at a new downstream focus F4.
The extension isolates the shower source from the detector, allowing
better shielding for $\gamma$-rays from the stopped positrons.

Simulation was done using a Monte Carlo beam transport program, 
REVMOC \cite{revmoc}, which calculates up to second-order optics
including the effects of multiple Coulomb scattering. The calculation was
done at the initial pion momentum of 76.8 MeV/c, which was degraded
to 73.6 MeV/c by a 2.0 mm thick Lucite
absorber. Separation of  53 mm  between the pion and positron images
was expected at F3. The positron image size was expected to be
16 mm (FWHM) with a small tail at the low momentum side, while
the pion size was 18 mm (FWHM) with a 10 \% tail.
Two thirds of the pions are lost due to decay-in-flight,
scattering, and collimation, resulting in a pion
to positron ratio of 100 expected
 at F4. Third order aberrations might cause some
broadening of the beam spot but were not expected to impact the suppression
factor.
\\

\subsection{The Extension}

The new extension starts at  F3 (0.9 m downstream of Q7), and
consists of a --70$^{\circ}$ bending magnet (B3)
at 1.5 m downstream of Q7, and a 30-cm diameter aperture
quadrupole triplet (Q8-Q10) after
B3. A 5 cm thick lead collimator with a 3 cm square hole placed at F3
stops the displaced positrons and redefines the pion image. 
The new focus F4 is 1.5 m downstream of Q10.
The total length of the extension between F3 and F4 is 4.5 m.
 To cause a momentum spread between positrons and pions,
1.45 mm and 2.0 mm thick Lucite degraders are mounted on the absorber wheel
near F1. Here the
momentum width of the incoming beam is restricted to 1.5 \% (FWHM)
by closing the SL1 horizontal slit to 1.5 cm. 
The beam after F1 is tuned for the pion momentum corresponding to
the pion energy after the energy loss. 
\\

\section{Measurements}

\subsection{Detector and measurements}

Tests were carried out in two stages; the first stage was done at F3
 before the installation of the extension,
 and the second at F4 after installation.
The incoming beam was measured at F3 or F4 with a telescope consisting of
two plastic scintillator beam counters  
(0.6 cm $\times$ 15.2 cm $\times$ 20.3 cm and 
0.3 cm $\times$ 3 cm $\times$ 4 cm), 
6 layers of 10-cm diameter wire chambers arranged
in the wire orientation of X-U-V-X-U-V (0$^{\circ}$ and $\pm 60^{\circ}$
with respect to the vertical axis), and a 48 cm diameter
 48 cm long NaI(T$\ell$)
crystal \cite{bnl} surrounded by two cylindrical layers of 8.5 cm thick,
 2$\times$25 cm long
pure CsI crystals \cite{e949}. The time of arrival of particles at the
beam counter with respect to
the cyclotron RF time provided particle identification
based on the time-of-flight (TOF)
together with the energy losses in the telescope counters.
\\

\subsection{Measurements without the Extension}

The pion and positron rates at 75 MeV/c were measured with the width of
 the momentum-defining horizontal slit SL1 set at 1.5 cm
to be 0.2 M/s and 0.05 M/s,
respectively,
for a 100 $\mu$A proton beam on the 1 cm thick Be target.
The vertical slits at SL1 and SL2 were set at 2.5 cm and 4 cm, respectively,
restricting the beam rates to 1/4 of the maximum rate.
The horizontal and vertical beam spot sizes at F3 were measured to be 
4.5 mm $\times$ 2.2 mm (rms), respectively, 
for positrons (10 \% wider for pions).

In order to suppress positrons based on the displacement at F3,
the tail in the beam profile needs to be minimized.
Effects of slits on the beam profile were studied.
Acceptance-limiting jaws SL0 upstream of the B1 magnet
 caused a broad tail up to 1/3 of the total positron intensity
 when they were closed to narrow (1 cm), while no tail was observed when SL0 
was wide open (12 cm).
 The momentum defining slits SL1 at F1 did not have an impact on the tail.
These observations indicated that
the jaws SL0 produced an additional source image 
that allowed different momentum components to pass through SL1.

\begin{figure}[htb]
\centering
\includegraphics*[width=8cm]{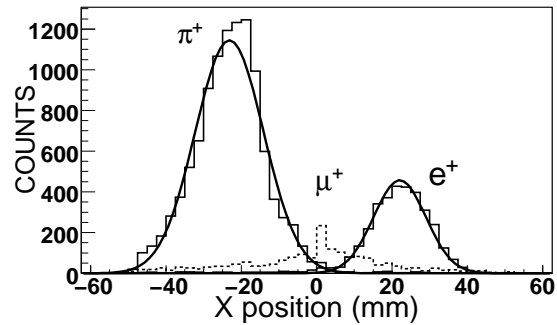}
\caption{Pion, muon and positron position distributions at F3 (histograms).
The heavy lines are fitted Gaussian curves for pions and positrons.}
\label{separation}
\end{figure}

When a 1.45 mm thick Lucite
absorber was inserted into the beam near F1,
the positron beam position at F3 was displaced by 46 mm 
with respect to the pion beam as shown in Fig.\ref{separation}.
In order to display pions and positrons (solid histograms) in the same plot,
the upstream momentum was raised only by 1 \%, and
the measured position and intensity might slightly 
be biased due to the geometry of the wire chambers.
The horizontal spot sizes of positrons and pions increased
 to 7.0 and 8.9 mm (rms), respectively, with a tail in the low momentum side.
For a thicker absorber (2.0 mm thick Lucite), the separation increased
to 50 mm.
\\

\subsection{Measurements with the Extension}

After initial tuning of the entire beam channel at 77 MeV/c
with a 5 cm thick lead collimator with a horizontal opening of 3 cm
placed at F3,
the 1.45 mm thick absorber was inserted and the downstream beam momentum
(only B2 and B3) was scaled to measure the pion and positron yields at F4.
\\

\begin{figure}[htb]
\centering
\includegraphics*[width=8cm]{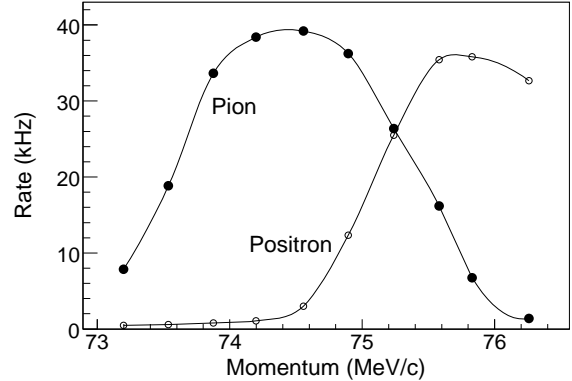}
\caption{Pion (closed circles) and positron (open circles) yields at F4 
for various downstream momenta.}
\label{epratio}
\end{figure}

Fig.\ref{epratio} shows a plot of pion and positron rates
at F4 vs. the downstream channel momentum. The rates are normalized to
a 100 $\mu$A proton current.
 After further tuning of beam focusing at F1 and F3, 
the  e$^+$ beam contamination
with respect to the pion rate was reduced to 1/60.
The horizontal and vertical beam spot sizes were 12 mm and 9 mm (rms), 
respectively.
The observed pion rate of 40 k/s, being consistent with the prediction,
can be increased without significantly affecting the vertical image at F4
to more than 100 k/s by opening the vertical SL1 and SL2 slits, which were
set at 2.5 cm and 4 cm, respectively.
\\

\subsection{Beam Channel Calibration Source}

Due to uncertainties in the fringe fields of the bending magnets, it
is usually difficult to obtain an absolute beam momentum calibration with
accuracy. Two sources commonly used for calibration 
are the high-momentum edges of positrons from 
$\mu^+ \rightarrow \mbox{e}^+ \nu \overline{\nu}$ and surface muons
from $\pi^+ \rightarrow \mu^+ \nu$.
The decay $\pi^+ \rightarrow \mbox{e}^+ \nu$ 
from stopped $\pi^+$ in the production target could
provide a high energy source with a definitive peak at 69.8 MeV/c.
Unlike the surface muon, the measurement of 
the decay $\pi^+ \rightarrow \mbox{e}^+ \nu$ from the production target
 does not require special equipment.

The major source of beam positrons above 55 MeV/c produced at 500 MeV
is from the decay
$\pi^0 \rightarrow \gamma \gamma$ promptly followed by
$\gamma$-ray conversion to electron-positron pairs in the
target material.
The positron momentum distribution is nearly flat
as shown in Ref.\cite{oldm13}.
These positrons are prompt with respect to the proton beam burst.

\begin{figure}[htb]
\centering
\includegraphics*[width=8cm]{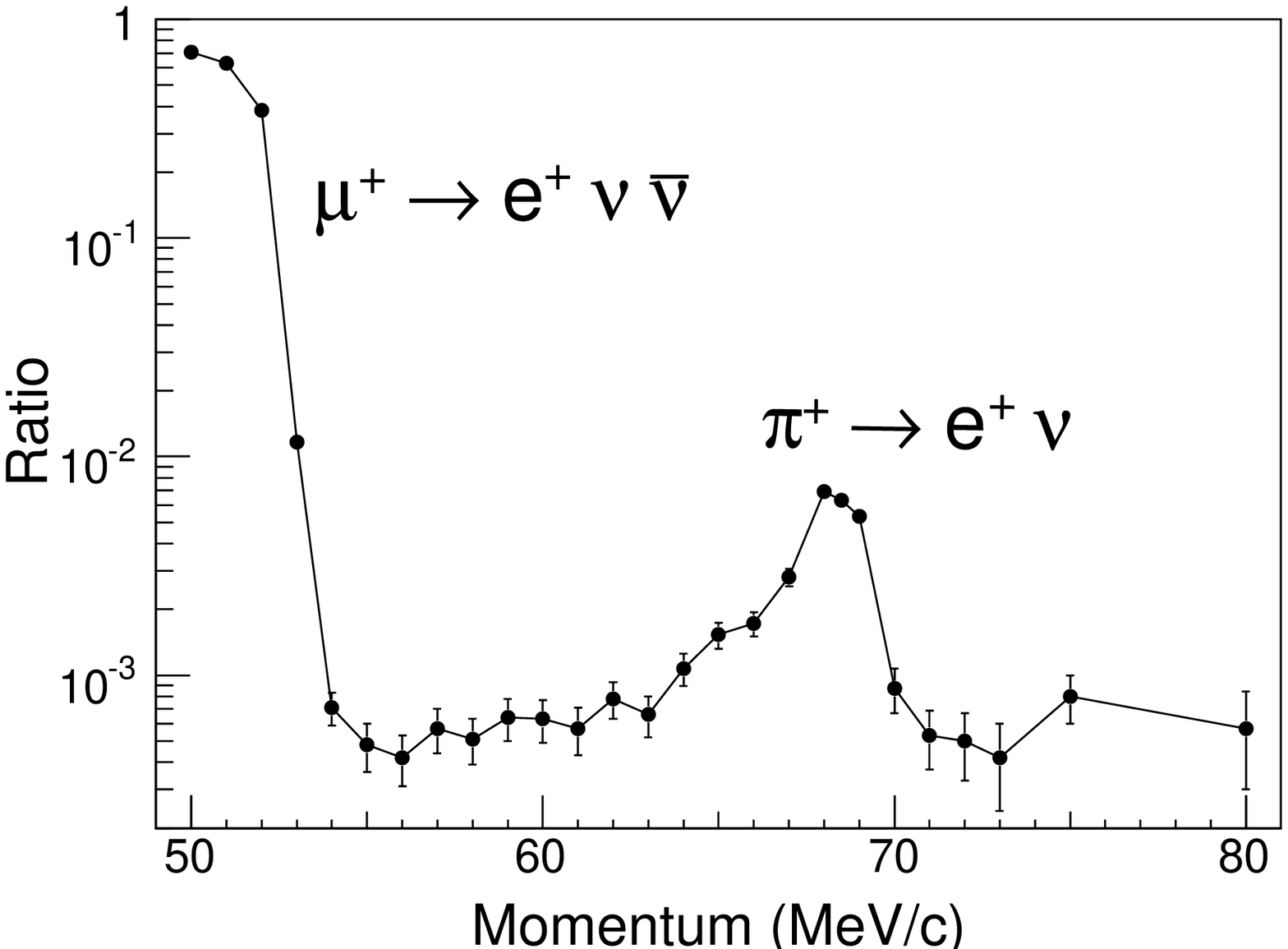}
\includegraphics*[width=8cm]{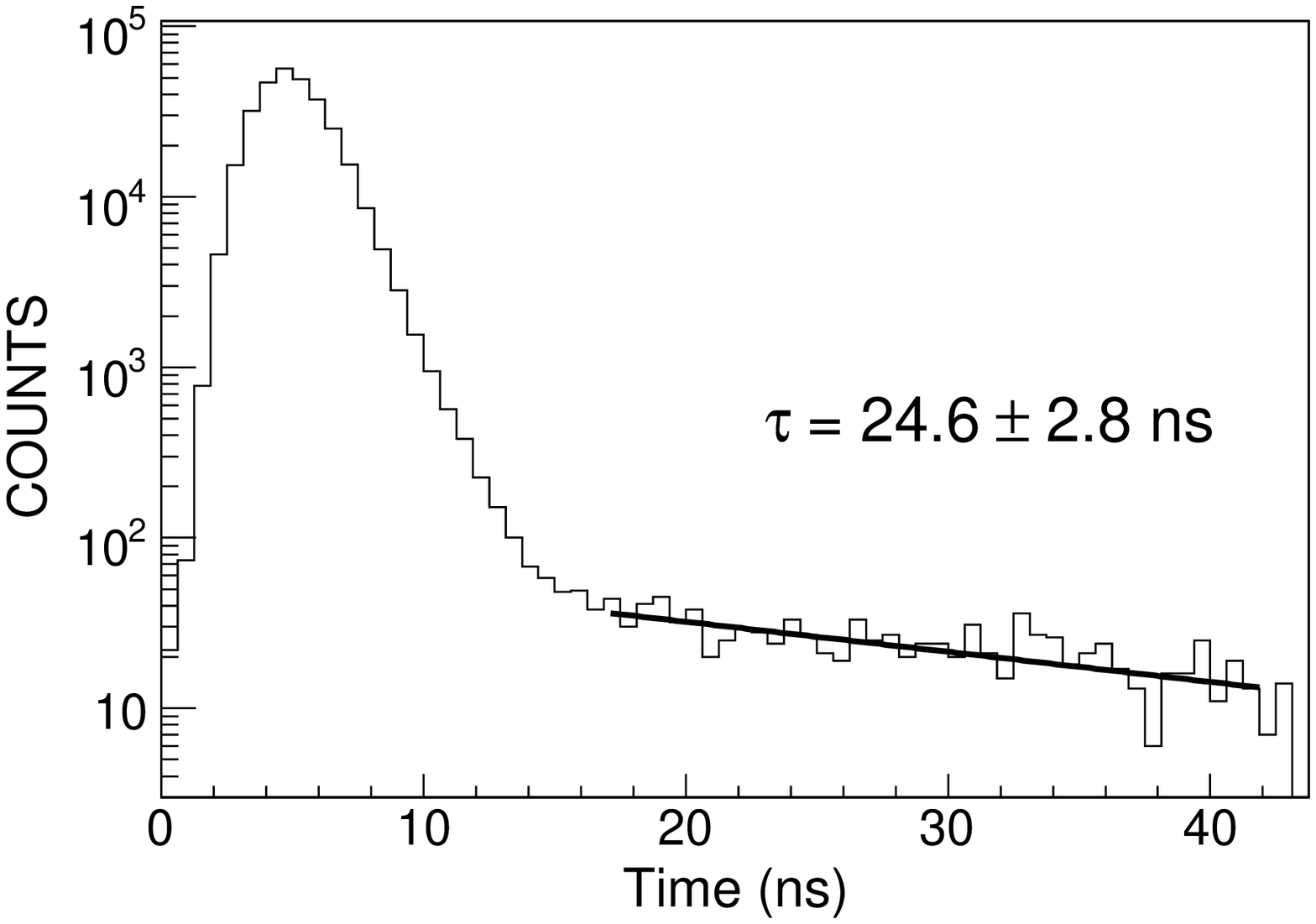}
\caption{(Top:a) Fraction of delayed positrons with the beam momentum and 
(Bottom:b) time
spectrum of 68.5 MeV/c positrons with respect to the proton burst. 
The fit of the delayed component to an exponential curve
is shown in the bold line.}
\label{pienu}
\end{figure}

We searched for delayed positrons coming from the beam line by varying
the beam channel momentum
between 50 MeV/c (below the $\mu^+ \rightarrow \mbox{e}^+ \nu \overline{\nu}$
 edge) and 80 MeV/c.
Tight cuts on energy loss in the beam counters suppressed
the pion and muon contaminations to a negligible level.
By selecting events with the beam energy in the NaI detector, pions and muons
were further suppressed as well as the
background from $\pi^+ \rightarrow \mu^+ \rightarrow \mbox{e}^+$ decays
coming from stopped pions near the detector.
At this point, events originated from $\pi^0$ were dominant. 
The  $\pi^+ \rightarrow \mbox{e}^+ \nu$ component was enhanced by
selecting delayed events using the TOF.
 Fig.\ref{pienu}a shows yields of delayed positrons
normalized to the total positrons with the beam channel momentum.
The edge around 52 MeV/c
 is from $\mu^+ \rightarrow \mbox{e}^+ \nu \overline{\nu}$ decays
from the production target.
The rate of delayed $\pi^+ \rightarrow \mbox{e}^+ \nu$ events was 0.6 \% of the
total positrons at the same beam momentum.
The peak momentum shift of 1.3 MeV was consistent with the energy loss of
positrons in the production target (0.5--1 MeV) and the uncertainty in
the calibration of 1 \%.
Fig.\ref{pienu}b shows a time spectrum for 68.5 MeV/c
positrons with respect to the proton burst (the RF signal). 
The delayed component has a decay constant of
24.6$\pm$2.8 ns, which is consistent with the pion 
lifetime \cite{pionlife}.
It is worthwhile to mention that these positrons are 
expected to be $\sim$100 \% 
circular-polarized.

By flipping the beam channel polarity to negative,
we also searched for delayed 50 MeV/c electrons from 
$\mu^- \rightarrow \mbox{e}^- \nu \overline{\nu}$
decays in the production target
as a potential high-intensity source of stopped negative muons.
The ratio of delayed and prompt electrons was measured to be
$(3.5 \pm 0.4) \times 10^{-3}$,
which is consistent with an estimate of $6 \times 10^{-3}$
based on the product of
the ratio of delayed and prompt positrons (2.9), 
the $\pi^- / \pi^+$ production ratio in this energy region (1/5), 
and the fraction of decay-in-flight of pions in which muons stop
in the target (1 \%) \cite{dif}. 
\\

\medskip
\section{Conclusion}

The TRIUMF M13 channel was modified to achieve a pion/positron ratio of
$>$50 using a differential energy-loss method.
The  performance of the extension satisfies the requirements of the PIENU
experiment, including a pion
rate of 100 k/s and positron suppression.
A new calibration source from $\pi^+ \rightarrow \mbox{e}^+ \nu$ decays in the
target  was identified for low-momentum beam channels.
\\

\section{Acknowledgment}

The authors wish to thank C. Ballard, N. Khan, R. Kokke,
 D. Evans, K. Reiniger, and
 the beam line group for the design and installation work.
This work was supported
 by the Natural Science and Engineering Research Council
and the National Research Council of Canada.
One of the authors (MB) has been supported by
US National Science Foundation grant Phy-0553611.
\\


\begin{thebibliography}{00}

\bibitem{triumf}  D.I.Britton $et~al.$, Phys. Rev. Lett. {\bf 68} (1992) 3000
and Phys. Rev. {\bf D49} (1994) 28.
\bibitem{psi}  G. Czapek $et~al.$, Phys. Rev. Lett. {\bf 70} (1993) 17.
\bibitem{pienu} TRIUMF proposal S1072, 2005.
\bibitem{oldm13} C.J. Oram $et~al.$,  Nucl. Instr. Meth. {\bf 179} (1981) 95.
\bibitem{meyer} D.I. Meyer, M.L. Perl and D.A. Glaster, Phys. Rev. {\bf 107}
(1957) 279.
\bibitem{revmoc} C. Kost and P. Reeve, TRIUMF report, TR-DN-82-28, 1982.
\bibitem{bnl} G. Blanpied $et~al.$, Phys. Rev. Lett. {\bf 76} (1996) 1023.
\bibitem{e949} I-H. Chiang $et~al.$, IEEE {\bf NS-42} (1995) 394.
\bibitem{pionlife} T. Numao $et~al.$, Phys. Rev. {\bf D52} (1995) 4855 and
V.P. Koptev $et~al.$, JETP Lett. {\bf 61} (1995) 877.
\bibitem{dif} T. Numao $et~al.$, Phys. Rev. {\bf D73} (2006) 092004.

\end{thebibliography}
\end{document}